\documentclass[fleqn,usenatbib,useAMS]{rasti}

\usepackage{newtxtext,newtxmath}

\usepackage[T1]{fontenc}

\DeclareRobustCommand{\VAN}[3]{#2}
\let\VANthebibliography\thebibliography
\def\thebibliography{\DeclareRobustCommand{\VAN}[3]{##3}\VANthebibliography}

\usepackage{graphicx}	
\usepackage{amsmath}	

\title[ASTRA's cosmic web]{Cosmic Web Classification through Stochastic Topological Ranking}

\author[J.E. Forero-Romero et al.]{
Jaime E. Forero-Romero $^{1,2}$\thanks{E-mail: je.forero@uniandes.edu.co},
Alejandro Palomino $^{1}$,
Felipe L. Gómez-Cortés$^{1}$,
Xiao-Dong Li$^{3,4,5}$
\\
$^{1}$Departamento de Física, Universidad de los Andes, Cra. 1 No. 18A-10, Edificio Ip, CP 111711, Bogotá, Colombia\\
$^{2}$Observatorio Astronómico, Universidad de los Andes, Cra. 1 No. 18A-10, Edificio H, CP 111711 Bogotá, Colombia\\
$^{3}$School of Physics and Astronomy, Sun Yat-Sen University, Zhuhai 519082, China\\ 
$^{4}$Peng Cheng Laboratory, No. 2, Xingke 1st Street, Shenzhen 518000, China\\
$^{5}$CSST Science Center for the Guangdong–Hong Kong–Macau Greater Bay Area, SYSU, Zhuhai 519082, China\\
}

\date{Accepted XXX. Received YYY; in original form ZZZ}

\pubyear{2015}

\begin{document}
\label{firstpage}
\pagerange{\pageref{firstpage}--\pageref{lastpage}}
\maketitle

\begin{abstract}
This paper introduces ASTRA (Algorithm for Stochastic Topological RAnking), a new method for classifying galaxies into cosmic web structures—voids, sheets, filaments, and knots—specifically designed for large spectroscopic surveys. 
ASTRA operates on observed galaxy positions and a corresponding random catalog, generating probabilistic cosmic web classifications for both datasets. 
The method's key innovation lies in using random points to trace underdense regions, enabling robust identification of cosmic voids that are poorly sampled by galaxies. 
We evaluate ASTRA using N-body simulations (dark matter-only and hydrodynamical) and SDSS observational data, performing both visual inspections and quantitative analyses of mass and volume distributions. 
The algorithm successfully produces void catalogs with size functions following theoretical expectations and demonstrates consistent environmental statistics across diverse datasets. 
Comparative analysis against established cosmic web classifiers confirms ASTRA's effectiveness, particularly for filament identification. 
By incorporating both observed and random points in its classification scheme, ASTRA provides a full cosmic web characterization without requiring density field interpolation or fixed geometric assumptions. 
The method's ability to quantify spatial correlations among different cosmic web components offers promising avenues for enhancing cosmological parameter constraints through non-standard clustering statistics.
\end{abstract}




\section{Introduction}
The large-scale distribution of galaxies in the Universe, known as the cosmic web, resembles a mesh of filaments spanning tens to hundreds of megaparsecs \citep{Bond}.
The primary driver behind the evolution of the cosmic web is widely acknowledged to be the anisotropic process of gravitational collapse \citep{Zeldovich}. 
Small perturbations in the primordial density field evolve, giving rise to elongated filaments as they collapse along their intermediate and shorter axes \citep{Springel}. 
As such, the cosmic web holds important information about the distribution of matter-energy in the Universe and the underlying laws governing its evolution.

Accurately characterizing the cosmic web could serve as a powerful tool for probing both dark matter and dark energy within the standard cosmological model, as well as potentially shedding light on the nature of gravity. 
However, a full understanding of the cosmic web requires not only the identification of overdense structures but also the characterization of underdense regions, which are poorly sampled by galaxies.

Due to the importance of cosmic web characterization, a variety of methods with distinct foundations have emerged in recent decades. 
Currently, classification methods for the cosmic web can be broadly categorized into five distinct classes:

\begin{enumerate}
\item Methods Based on the Hessian (Geometric and Multiscale): This category involves methods relying on the calculation of the Hessian of density, gravitational potential, or velocity field \citep{T-web,V-web}. Implementing Hessian-based techniques requires spatial coordinates interpolation and the creation of a continuous field, typically achieved using Fourier transforms and smoothed with a Gaussian kernel. 
Geometric methods within this category establish connections between the morphology of density fields and the classification of points into cosmic structures. 
In contrast, multiscale methods use different scales when smoothing the field over which the Hessian is computed.
The \textit{Multiscale Morphology Filter} algorithm by \citep{MMF-2} is a prime example employing multiscale techniques.

\item Graph-Based Methods: Historically significant, these methods utilize graph theory principles to analyze matter distribution. 
The Minimum Spanning Tree (MST) algorithm is a notable example as it was one of the earliest approaches for studying filamentary structures \citep{Barrow, MST}. 
The MST is a tree-like structure that connects all the nodes of a given graph with the minimum total edge weight possible, finding the shortest path that connects all the nodes without forming any cycles. 
More recent examples include the $\beta$-skeleton \citep{2019MNRAS.485.5276F}, a geometric graph where edges connect points if the distance between them is not greater than a certain proportion (parameterized by $\beta$) of the distance to their nearest neighbor. 
This graph has been used to describe the cosmic web \citep{2021ApJ...922..204S} and it is starting to be used to constrain cosmological parameters \citep{ImprovingSDSS}.

\item Stochastic Methods: Stochastic methods use statistical evolution of results derived from graphical or geometric concepts. 
An example is the approach outlined by \citep{bisous}, which classifies galaxies through hypothetical high-density cylinders with varying geometric configurations in each iteration. 
Stochastic methods offer an advantage in offsetting errors caused by data peculiarities and, like the method presented in this study, they do not require calculating a density field, utilizing only the coordinates of each point in the catalogs. 
However, this method is limited to finding filamentary structures.

\item Topological Methods: Similar to Hessian-based methods, topological methods aim to find general features in the morphology of filamentary structures. 
While Hessian-based methods focus on local search within geometric structures, topological methods study galaxy connections based on topological approaches. 
Two prime examples are the Disperse \citep{disperse} and the SpineWeb method \citep{spineweb}, which use Morse theory and Delaunay tessellations to study the cosmic web. 
Another example is void finders that use Voronoi tessellations together with linking techniques to define aspherical voids \citep{VIDE}.

\item Phase Space Methods: Phase space methods prioritize studying the dynamics of structure formation by analyzing the evolution of Phase Space generated initial conditions in simulations to a given epoch. 
An example is the method defined in \citep{Origami}, which classifies points based on features found by studying the Lagrangian of the 6-dimensional phase space system. 
This method is limited to simulations where the positions and velocities of simulation particles are well determined both in the initial conditions and at another timestep of interest.
\end{enumerate}

In this study, we introduce a new method called ASTRA (Algorithm for Stochastic Topological Ranking) designed for the classification of galaxies within the Cosmic Web. 
The main goal of our methodology is to provide a complete view of the cosmic web, including both overdense and underdense regions. 
We achieve this by including random points in the classification scheme.

ASTRA innovates in three key aspects:
\begin{enumerate}
    \item It operates on the typical data structure derived from large scale structure catalogs from spectroscopic surveys, namely Cartesian coordinates computed from angular positions and redshifts and their corresponding random data distribution.
    \item It eliminates the need for interpolation, smoothing, or imposing fixed geometrical shapes on the data.
    \item It has the capability to identify all four cosmic web types: voids, sheets, filaments, and knots.
\end{enumerate}

By incorporating both data and random points in its classification scheme, ASTRA provides a unique and complete view of the cosmic web structure. 
This approach allows for the identification of underdense regions (voids) that are typically challenging to characterize due to the lack of galaxies in these areas.

We structure this ASTRA presentation paper as follows. 
In \S \ref{sec:methodology} we describe the algorithm and the different applications of the ASTRA outputs, with particular emphasis on how the inclusion of random points enables the identification of underdense regions. 
In \S \ref{sec:quantifying} we describe the statistics and datasets that we are going to use to quantify ASTRA's performance, including its ability to characterize both overdense and underdense structures. 
We continue in \S \ref{sec:results} with the results together with a general discussion on ASTRA's performance and capabilities, highlighting its complete view of the cosmic web. 
Finally, we present our conclusions in the last section.

\section{Methodology}
\label{sec:methodology}

\begin{figure*}
	\includegraphics[width=1.9\columnwidth]{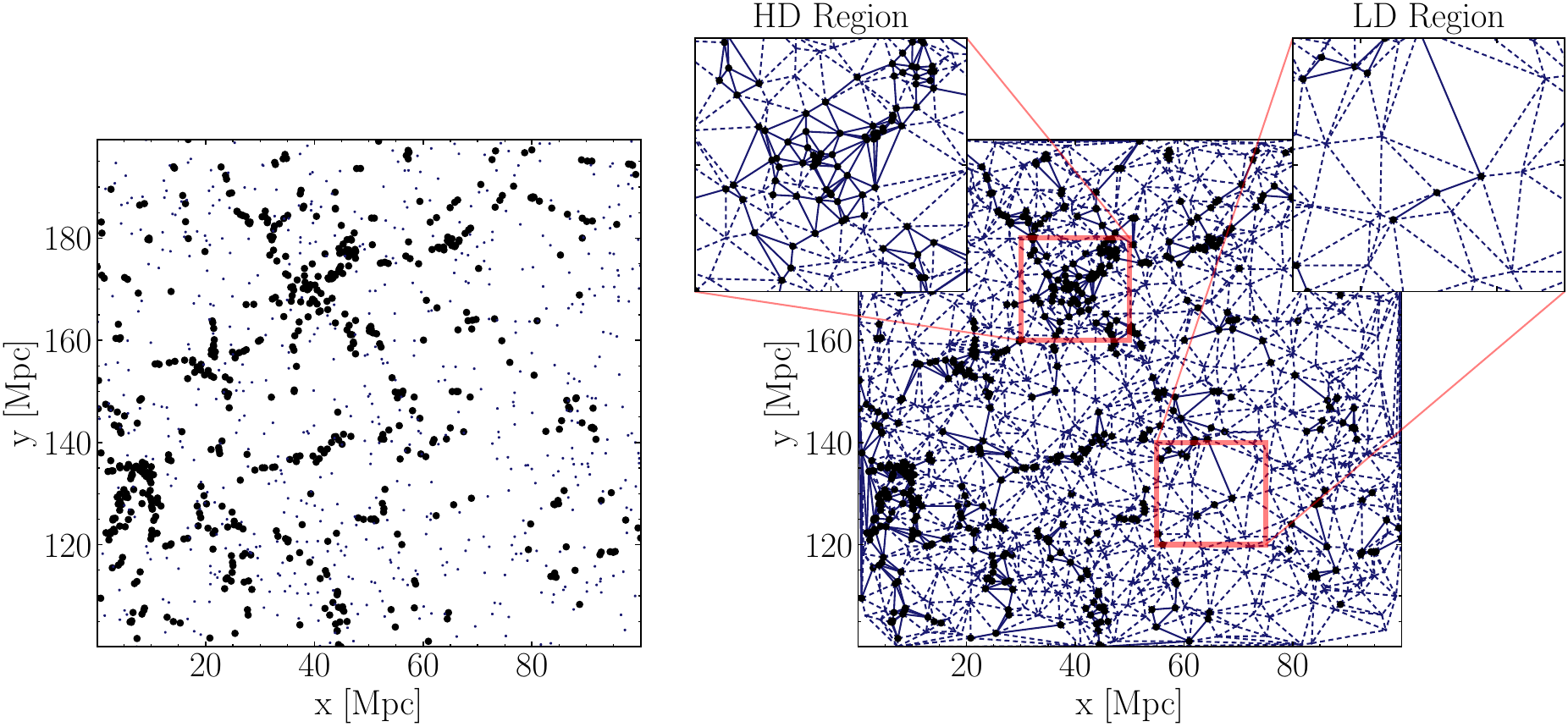}
    \caption{Illustration of the ASTRA method applied to a 2-dimensional dataset of 50 points. 
    The left panel displays the input object points (larger black points) and the randomly generated points uniformly distributed in the Monte Carlo iteration (blue points, smaller than the object points). 
    The right panel depicts the Delaunay graph, with solid lines connecting two object points and dashed lines connecting at least one random point. 
    In addition, this graph has 2 zoom-ins (of areas with high and low density from left to right), which qualitatively show the difference between the proportion of the number of connections of the object points between them over the connections with random points. 
    In the HD region solid lines (connections to object points) dominate, while in the LD region dashed lines (connections to random points) dominate.}
    \label{pathsALL}
\end{figure*}

The ASTRA method is a stochastic algorithm that classifies points in 3D space into one of four classes: voids, sheets, filaments, and knots. 
This classification is based on local computations made from a graph.
The algorithm also needs as an input a random catalog of points that follows the number density distribution of the object points, hence its stochasticity.

\subsection{Algorithm Description}

ASTRA takes as input two datasets:

\begin{enumerate}
    \item $\mathcal{O} = \{o_1, o_2, ..., o_{N_O}\}$: The set of observed or simulated objects, where each $o_i$ represents a galaxy or dark matter halo with known 3D coordinates. 
    We refer to this as the "object catalog".
    \item $\mathcal{R} = \{r_1, r_2, ..., r_{N_R}\}$: The set of randomly distributed points that follow the same selection function and geometric constraints as $\mathcal{O}$. 
    We refer to this as the "random catalog". 
    It represents an unclustered Poisson sampling of a constant background density. 
\end{enumerate}

The random catalog follows the same selection function and geometrical constraints as the object catalog, meaning it mimics the survey's observational limitations and data processing cuts. 
For observational data, this includes reproducing the angular selection boundaries (such as the declination and right ascension ranges), the radial distance distribution derived from the redshift selection cuts, and any magnitude or completeness limits applied to the galaxy sample. 
For simulation data, the random points are uniformly distributed within the same volume boundaries as the simulated objects. 
This approach follows standard practices in large-scale structure analysis, where random catalogs are essential for accurate correlation function measurements and statistical analyzes \citep{1993ApJ...412...64L,1993ApJ...417...19H}.

In our implementation, we generate an equal number of random points as objects, i.e. $N_R = N_O$, to ensure a consistent local mean point density in both sets. 
This also requires that the volume spanned by $\mathcal{O}$ and $\mathcal{R}$ remains the same.

The merged catalog $\mathcal{M} = \mathcal{O} \cup \mathcal{R}$, comprising both the object and the random catalogs, undergoes Delaunay triangulation. For each point $p_i \in \mathcal{M}$, we compute the following:

\begin{itemize}
    \item $N_\mathcal{O}(p_i)$: The number of connections to points in $\mathcal{O}$
    \item $N_\mathcal{R}(p_i)$: The number of connections to points in $\mathcal{R}$
\end{itemize}

Using these quantities, we calculate the dimensionless parameter $r$ for each point $p_i$ in the merged catalog:

\begin{equation}
r(p_i) = \frac{N_\mathcal{O}(p_i) - N_\mathcal{R}(p_i)}{N_\mathcal{O}(p_i) + N_\mathcal{R}(p_i)}.
\end{equation}

Positive $r$ values indicate a greater number of connections to object points, while negative values indicate a higher number of connections to random points. 
Based on the value of $r$, each point in the merged catalog is classified into a cosmic web type according to predefined threshold values outlined in Table \ref{c1}.

We adopt this formulation over an overdensity-based approach, such as $N_\mathcal{O}(p_i)/N_\mathcal{R}(p_i) - 1$, as it allows us to handle situations with high object point density where no random points, $N_\mathcal{R}(p_i)=0$, are linked in the Delaunay triangulation.

The classification process can be iterated $N_{\text{iter}}$ times, with the random point distribution changing at each iteration, resulting in $N_{\text{iter}}$ distinct classifications for each object point. 
For these object points, the algorithm estimates the probability $p_w$ of being classified into one of the four web elements (knot, filament, sheet, or void) as the ratio of the number of times the point was classified into each of these elements to the total number of iterations $N_{\text{iter}}$.

Figure \ref{pathsALL} illustrates the ASTRA methodology applied to a two-dimensional example with 50 points. 
The left panel shows the input object points (black) and randomly generated points (blue, smaller), while the right panel displays the resulting Delaunay graph with connections between points. 
The zoomed insets demonstrate how the proportion of connections varies between high-density (HD) regions (where solid lines connecting object points dominate) and low-density (LD) regions (where dashed lines connecting to random points are more prevalent). 
This connectivity pattern forms the basis for our cosmic web classification scheme.

\subsection{Choice of Graph Method}

In ASTRA we choose Delaunay tessellation as our primary graph-based method for analyzing the cosmic web structure. This decision is based on several key factors:

\begin{enumerate}
    \item Simplicity of implementation: Delaunay tessellation provides a straightforward and computationally efficient approach to constructing a graph representation of the cosmic web. 
    Its well-established algorithms are easily implemented and optimized for large-scale structure analysis.
    
    \item Broad usage in the community: Delaunay tessellation has been widely applied in cosmology and astrophysics. 
    Its prevalence facilitates comparisons with other studies and integration into existing scientific workflows.
    
    \item Versatility: While our primary focus is classifying cosmic web elements, Delaunay tessellation offers a versatile foundation for various analyses, including void finding, filament detection, and density estimation.
\end{enumerate}

Furthermore, the use of Delaunay tessellations for cosmic web analysis has a rich history in the field. 
The Delaunay Tessellation Field Estimator (DTFE) developed by \citet{2000A&A...363L..29S,2009LNP...665..291V} established the foundation for exploiting the geometrically adaptive properties of Delaunay tessellations in cosmological applications. 
DTFE has demonstrated exceptional performance in reconstructing density and velocity fields from discrete point distributions while preserving both the multiscale character and local geometry of cosmic structures \citep{2011arXiv1105.0370C}.

The DTFE methodology has subsequently been incorporated into several cosmic web classification algorithms. 
The Multiscale Morphology Filter (MMF) and its successor NEXUS \citep{MMF-2,NEXUS} utilize DTFE-reconstructed density fields as the basis for their scale-space analysis of cosmic web morphology. 
Similarly, the Watershed Void Finder (WVF) \citep{2007MNRAS.380..551P} employs DTFE density reconstructions in conjunction with watershed techniques for void identification. 
The topological analysis framework DisPerSE also builds upon DTFE density field reconstructions \citep{disperse}.

While ASTRA shares with these methods the fundamental insight that Delaunay tessellations provide an optimal framework for cosmic web analysis, our approach differs in its direct use of the tessellation connectivity without requiring density field reconstruction or interpolation. 
Rather than first estimating a continuous density field and then applying morphological filters, ASTRA operates directly on the discrete connectivity patterns of the Delaunay graph, combined with the novel inclusion of random points to characterize both overdense and underdense regions.

As mentioned in the introduction, researchers have explored other graph-based methods for cosmic web analysis. 
Notable examples include the Minimum Spanning Tree (MST) algorithm and, more recently, the $\beta$-skeleton method, which has gained attention as a tool for cosmic web analysis \citep{2019MNRAS.485.5276F,2021ApJ...922..204S,ImprovingSDSS}.

In the context of void finding, an important aspect of cosmic web classification, we previously explored the $\beta$-skeleton using the same object-random framework employed in this study. 
However, our investigations did not reveal significant differences between the $\beta$-skeleton approach and the Delaunay tessellation method presented here, particularly in identifying and characterizing cosmic voids \citep{gomez2019beta}.

Given the comparable performance and the aforementioned advantages of Delaunay tessellation, we have chosen this method for ASTRA. 
Nevertheless, we acknowledge the potential of alternative graph-based methods and encourage further comparative studies to refine and optimize cosmic web classification techniques.

\subsection{The importance of $N_R=N_O$}

In our definition of $r$, having a ratio of the number of random points to the number of object points, denoted as $\rho = N_R / N_O=1$, is important for the effectiveness and simplicity of the algorithm. 
This setup with $\rho=1$ simplifies the interpretation of the cosmic web classification: positive $r$ values clearly indicate overdensity (more connections to object points), while negative values indicate underdensity (more connections to random points).

By maintaining $\rho=1$, ASTRA establishes a consistent reference for the mean density across the entire volume. 
This consistency is important for comparing different regions and structures within the cosmic web. 
If $\rho \neq 1$, the formula for $r$ would need to be adjusted with scaling factors to account for the unequal number of random and object points. 
This would unnecessarily complicate the algorithm.

The choice of $\rho=1$ is particularly effective in identifying random points in underdense regions (voids) that are typically challenging to characterize due to the lack of galaxies in these areas. 
The equal number of random points ensures that these regions are well-sampled. 
This aspect of ASTRA is especially valuable, as it addresses a common limitation in cosmic web classification methods.
Furthermore, as shown in \cite{gomez2019beta}, using values of $\rho>1.6$ results in percolation of voids throughout the volume, that is, identified voids start to increase in size due to a percolation process that links the voids throughout the volume. 
For large values of $\rho\approx 2$ the percolation typically reaches a state where a single void has merged into itself all the random points classified as voids.

\subsection{Threshold Determination}

The classification thresholds listed in Table \ref{c1} are established through a statistical procedure designed to maintain robustness across varying density environments. 
The process for determining these thresholds is as follows.

Two synthetic point sets are generated within identical spatial boundaries and with equal number densities. 
One set emulates a homogeneous distribution, analogous to the observed data, while the other replicates the properties of the accompanying random catalog. 
This setup preserves the $\rho = 1$ ratio between data and random points, as previously discussed.

The ASTRA algorithm is then applied to both sets, computing $r$ values for each point. 
The distribution of these $r$ values is analyzed to identify symmetric thresholds that encompass 99\% of the data centered on $r = 0$. 
This yields cutoff values of $-0.9$ and $0.9$, corresponding to the lower and upper bounds, respectively.

Selecting the 99th percentile reflects a compromise between statistical rigor and practical classification needs. 
This threshold captures deviations of approximately 2.6$\sigma$ from the mean of the connectivity distribution, offering a stringent yet inclusive criterion for identifying genuine cosmic web features while suppressing spurious detections due to noise. 
A more restrictive choice (e.g., 99.9th percentile) would isolate only the most extreme structures, whereas a looser one (e.g., 95th percentile) would admit more features at the cost of increased contamination.

In the context of a statistically uniform distribution, these thresholds imply that voids and knots should not be detected, ensuring that ASTRA highlights only the most statistically significant fluctuations. 
This criterion effectively balances sensitivity to meaningful cosmic web structures with resistance to noise. 
The intermediate transition between sheets and filaments naturally arises from the sign of $r$, distinguishing underdense from overdense environments around $r = 0$.

Empirical tests confirm that the 99th percentile threshold yields classifications with appropriate mass segregation and volume-filling fractions. 
While somewhat subjective, like threshold choices in T-web, V-web, DisPerSE, or watershed-based methods, this selection performs well across a range of applications. 
Users may tailor the threshold to suit their scientific goals.

Although derived from idealized, homogeneous configurations, these thresholds remain effective for real data due to ASTRA’s local approach and the inherent randomness of the point distribution. 
The inclusion of random points, the key innovation of ASTRA, enables the detection of voids, which are otherwise difficult to characterize due to their sparsity in galaxy surveys.

\subsection{Classification uncertainty}

A probabilistic interpretation of ASTRA's cosmic web classification can be built through the usage of multiple random catalogs, which enables uncertainty quantification in the classification process. 

Using a large number of random catalogs, set to 100 in this paper, we produce the cosmic web classification using different random catalogs as an input, that is we do 100 different classifications, where the only change is in the random catalog,
This approach allows us to assign probabilities to each point's classification while simultaneously evaluating the confidence level of these assignments. 

For each point, the probability $p_w$ of belonging to a specific cosmic web environment is calculated as the fraction of classification iterations that result in that particular web type.

To quantify the uncertainty in these classifications, we employ the normalized information entropy function \citep{1949mtc..book.....S}:
\begin{equation}
H = -\frac{1}{\log_2{4}}\sum_{w=1}^4 p_w \log_2(p_w)
\label{eq:entropy}
\end{equation}
where $p_w$ represents the probability for a point to belong to each of the four cosmic web environments. 
This entropy measure ranges from 0 to 1, providing a quantitative metric for classification certainty. 
A value of $H = 0$ indicates complete confidence, occurring when a point receives the same classification across all random catalog iterations ($p_w = 1$ for one environment and 0 for all others). 
The maximum value of $H = 1$ occurs only when all probabilities are equal ($p_w = 0.25$ for all environments), indicating maximum classification uncertainty. 
Intermediate values, such as $H = 0.5$, typically arise when classification is split primarily between two environments. 
For example, a point with classification probabilities of [0.68, 0.31, 0.01, 0.00] has an entropy of 0.48, while a more confident classification of [0.95, 0.03, 0.01, 0.01] yields a lower entropy of 0.17.

\begin{table}
\caption{Classification of points according to the distribution of neighbors given by the Delaunay triangulation.}
\centering
\begin{tabular}{|c|c|}
\hline \hline
Condition   & Classification     \\ \hline
$-1 \leq r \leq -0.9 $   & void     \\ 
$-0.9 < r \leq 0 $   & sheet    \\ 
$0 < r \leq 0.9 $         & filament \\ 
$0.9 < r \leq 1 $ & knot  \\ \hline
\end{tabular}
\label{c1}
\end{table}

\section{Data}
\label{sec:quantifying}

We use three different datasets as input to test ASTRA. 
The first simulated dataset used in this study is from the Tracing the Cosmic Web (TCW) comparison project \citep{TCW}.
The simulation is a dark matter-only N-body simulation, with a box size of 200 Mpc $h^{-1}$ and 512$^3$ particles, using the Gadget-2 code \citep{spring} and cosmological parameters of $h = 0.68$, $\Omega_M = 0.31$, $\Omega_\Lambda = 0.69$, $n_s = 0.96$, and $\sigma_8 = 0.82$. 
A dataset of 281,465 dark matter haloes was obtained using a Friends-of-Friends (FOF) algorithm \citep{1985ApJ...292..371D} with a linking length of $b=0.2$ and a minimum of 20 particles for this analysis, resulting in a number density of $11.0 \times 10^{-3}$ Mpc$^{-3}$.
This simulation is advantageous as it has already been analyzed by 11 different methods, which have also classified each point into one of the four cosmic structures. 

The second catalog of simulated data used in this study is from the Illustris-TNG (TNG) project \citep{tngsim}, which includes simulations of dark matter from a redshift of $z=127$ to $z=0$ in boxes of sizes in boxes of comoving sizes 35 Mpc $h^{-1}$, 75 Mpc $h^{-1}$, and 225 Mpc $h^{-1}$ (corresponding to TNG50, TNG100, and TNG300, respectively).
The cosmological parameters used in these simulations are $h=0.6774$, $\Omega_{\Lambda}=0.6911$, $\Omega_M=0.3089$, $\Omega_B=0.0486$, $\sigma_8=0.8159$, $n_s=0.9667$, and $H_0= 100h$ km s$^{-1}$ Mpc$^{-1}$. 
For this study, we selected the TNG300-1 catalog, which has a box size of 225 Mpc $h^{-1}$, a redshift of $z=0$, and 2500$^3$ particles. 
A filter was applied to select only those galaxies with a stellar mass above a certain threshold (log${10}(M/M_{\odot} \ h^{-1}) > M_{\text{lim}}$, where $M_{\text{lim}} = 9$), resulting in a sample of 221,279 galaxies, corresponding to a number density of $8.2 \times 10^{-3}$ Mpc$^{-3}$. 
This limit corresponds to the lowest stellar mass limit we can reach at which the stellar mass function is considered to have been converged \citep{2018MNRAS.475..648P}.
We note that galaxies at this mass threshold are resolved with approximately 100 star particles in TNG300, which, while at the resolution limit for detailed galaxy formation studies, is sufficient for cosmic web analysis where the primary requirement is accurate spatial positioning of tracers.

The third catalog used in our study is an observational data catalog from the Sloan Digital Sky Survey (SDSS) Data Release 7 \citep{Abazajian_2009}. 
Specifically, we use data from the NYU Value-Added Galaxy Catalog \citep{NYU}, which includes large-scale structure samples constructed from SDSS data. 
Initially, we had 559,028 galaxies. 
To create a volume-limited sample, we applied cuts to the r-band magnitude and redshift. 
We selected all galaxies with an r-band magnitude of $M_r \leq -20$ and a redshift $z \leq 0.114$. 
We chose this magnitude limit to preserve all galaxies brighter than $L_*$ \citep{SDSSlimit}, as they reveal the structure of the cosmic web. 
Additionally, we selected galaxies in the declination range between 0° and 50°, and right ascension range between 120° and 230°, resulting in a final sample of 90,655 galaxies.
This corresponds to a total comoving volume of $5.2 \times 10^{7}$ Mpc$^{3}$ and a number density of $1.7\times 10^{-3}$ Mpc$^{-3}$
This specific angular range was chosen to simplify the process of generating random points. 
Then, we compute the Cartesian coordinates through a two-step process. 
First, we convert the redshift into a comoving radial distance using the cosmological parameters from the TNG simulation. 
Then, we utilize the angular coordinates (right ascension and declination) along with the radial coordinate to compute the final $x$, $y$, $z$ Cartesian coordinates. 
It is important to note that we do not apply corrections for redshift space distortions. 
This limitation may lead to an overclassification of galaxies as filaments, as high-density peaks are elongated along the line of sight in redshift space \citep{2019MNRAS.485.5276F}, potentially blending into neighboring filamentary regions.
It is important to note that neither of the simulation-based catalogs (TCW and TNG) include modeling of observational effects such as survey incompleteness or redshift space distortions.

It is important to acknowledge that our three datasets represent a heterogeneous collection with fundamentally different selection criteria and object types. 
The TCW catalog contains dark matter halos with a minimum mass threshold of approximately $2.0 \times 10^{12} M_{\odot}$ $h^{-1}$, the TNG catalog includes galaxies selected by stellar mass above $10^9 M_{\odot}$ $h^{-1}$, and the SDSS sample comprises galaxies selected by r-band luminosity ($M_r \leq -20$). 
These different selection functions result in varying number densities: TCW ($11.0 \times 10^{-3}$ Mpc$^{-3}$), TNG300 ($8.2 \times 10^{-3}$ Mpc$^{-3}$), and SDSS ($1.7 \times 10^{-3}$ Mpc$^{-3}$). 
While this heterogeneity demonstrates ASTRA's versatility in handling diverse data types, from dark matter halos to observed galaxies, it does introduce limitations in making direct quantitative comparisons between datasets. 
The different mass/luminosity thresholds and sampling strategies mean that each catalog traces different populations of objects and potentially different aspects of the underlying cosmic web structure. 
Nevertheless, the broad similarity in number densities between TNG and TCW, combined with the ability to analyze sparse observational data like SDSS, showcases the method's applicability across the range of datasets commonly used in large-scale structure studies.

\section{Results}
\label{sec:results}

In this section, we present a complete approach to quantify the outputs of ASTRA across six distinct aspects. 
Our analysis begins with a visual inspection of the different cosmic web types identified by the algorithm. 
We then proceed to examine the distributions of the $r$ values and employ classification entropy to assess the robustness of web type assignments. 
Following this, we quantify the mass, luminosity, and volume distributions across the various web types. 
To demonstrate ASTRA's practical applications, we illustrate its capability to produce catalogs of the void size function. 
Our analysis extends to spatial correlations, where we quantify relationships across web types using both auto and cross-correlation functions. 
Finally, we conduct a comparative analysis, evaluating ASTRA's results against different published methods for cosmic web classification. 
This approach allows us to thoroughly evaluate ASTRA's performance and validate its effectiveness in cosmic web analysis.

\begin{figure*}
	\includegraphics[width=2\columnwidth]{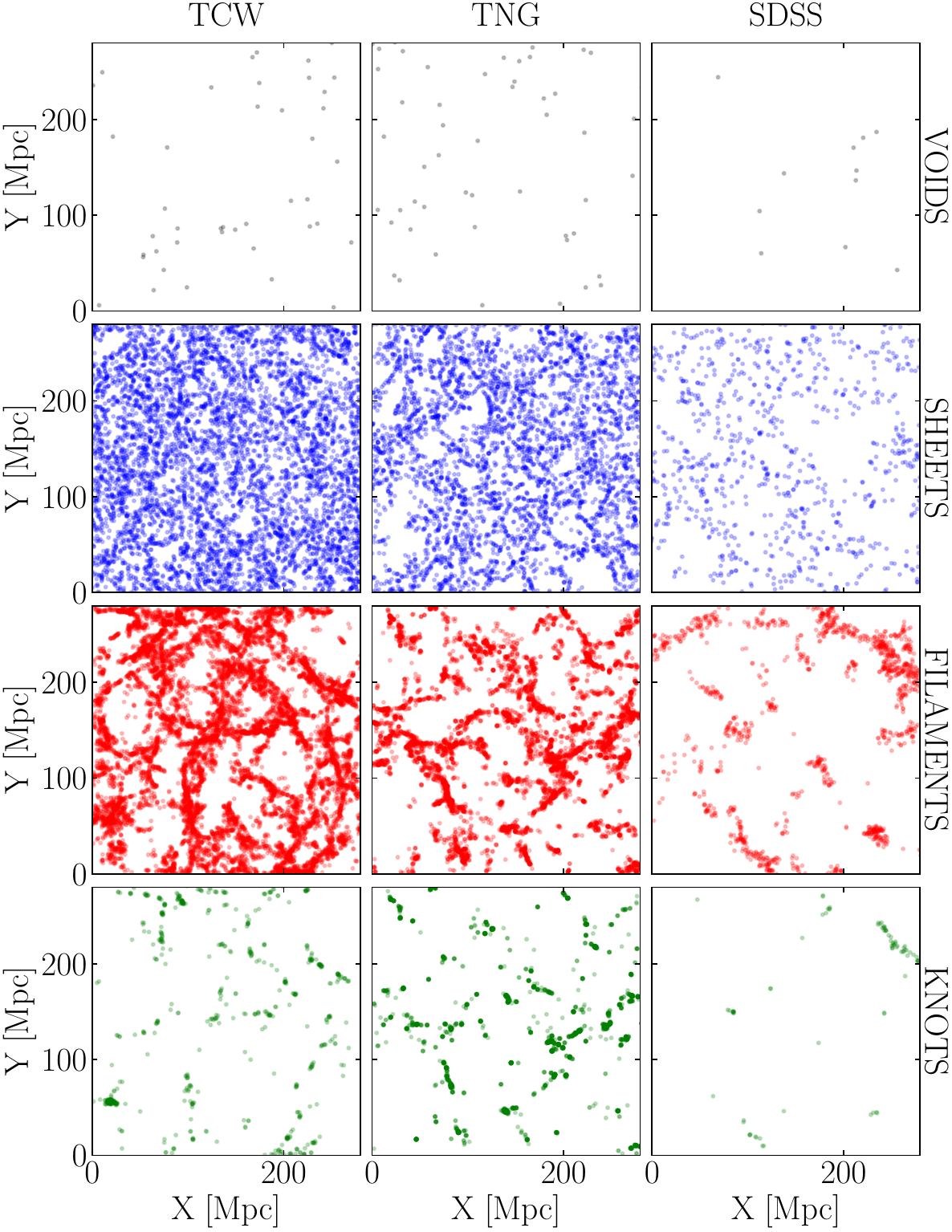}
    \caption{Plot of the four cosmic web types classification given by ASTRA for each of the three object catalogs. 
    The columns, from left to right, show the results obtained for TCW, TNG and SDSS, while the rows, from top to bottom, show the points belonging to each type organized in ascending order of density; voids, sheets, filaments and knots. 
    The image shows a z-slice that is 20 Mpc wide, 280 Mpc on each side.
    The structure of each of the painted points was chosen using a single Monte-Carlo iteration.
    In the case of the SDSS dataset the observer is located at the origin of the coordinate system.}
    \label{classesALL}
\end{figure*}

\begin{figure*}
	\includegraphics[width=2\columnwidth]{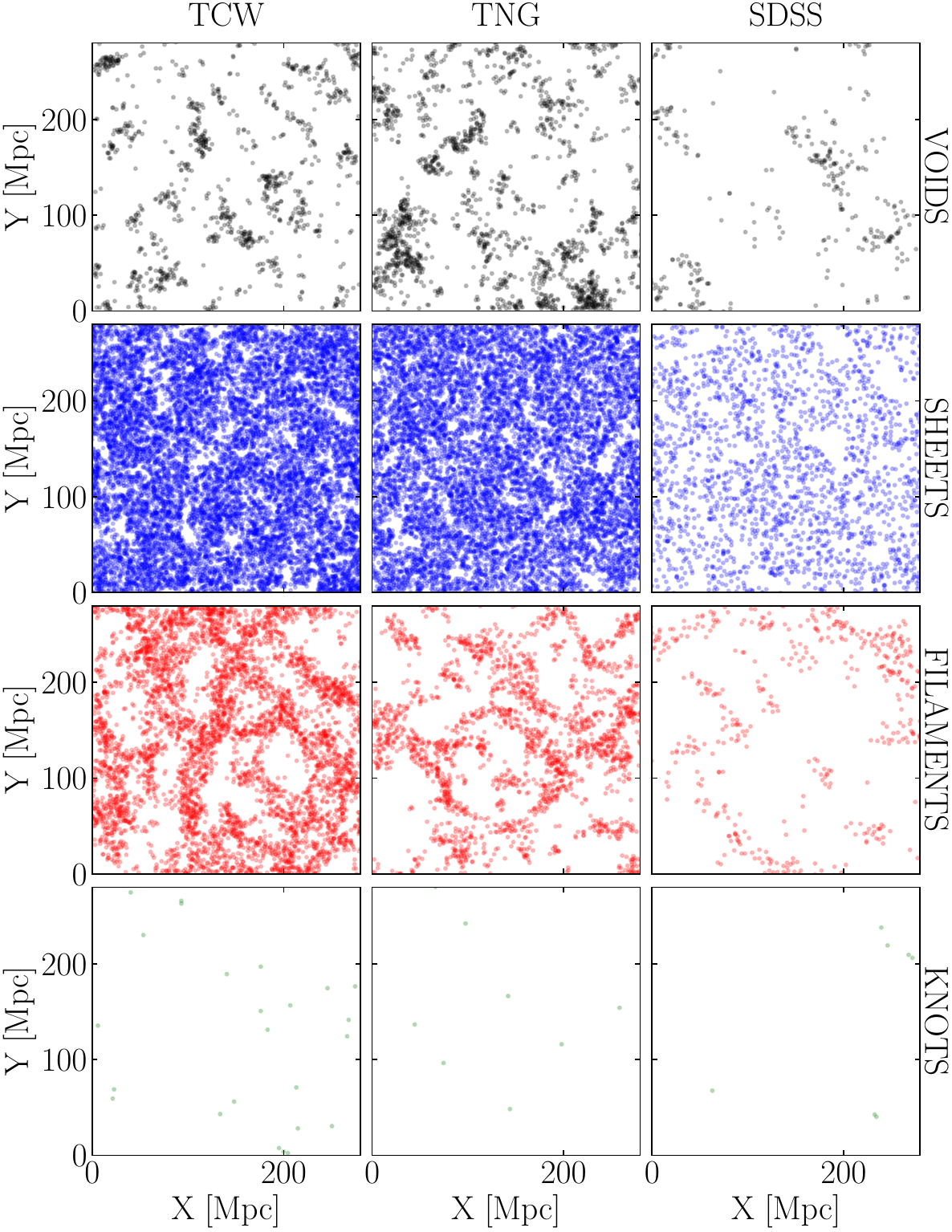}
    \caption{Plot of the classification given by ASTRA for the uniformly generated random points in each of the three data catalogs.
    following the same pattern as in Figure \ref{classesALL}. 
    The structure of each of the painted points was chosen with the results of the iteration classification. 
    The most relevant characteristic is that now voids are clearly visible by multiple random tracers.}
    \label{classesALLRandom}
\end{figure*}

\subsection{Visual inspection}

Since cosmic web algorithms are designed and tuned to various degrees to reproduce visual impressions \citep{TCW}, we begin with a qualitative exploration of these visual patterns before proceeding to quantitative analysis.

Figure \ref{classesALL} presents a complete view of the ASTRA algorithm's classification results for four cosmic web structures across three different catalogs: dark matter-only simulations (TCW), hydrodynamical simulations (TNG), and observations (SDSS).

This figure displays points labeled as objects, arranged from top to bottom in order of increasing average density: voids, sheets, filaments, and knots. 
Each visualized slice maintains consistent dimensions of 280 Mpc in width and 20 Mpc in depth, enabling direct comparison across all cases.

The results in Figure \ref{classesALL} demonstrate ASTRA's capability to produce expected patterns across various inputs:

\begin{itemize}
    \item Voids: Appear as sparse, low number-density regions, clearly representing the objects in underdense areas of the cosmic web.
    \item Sheets: Exhibit a nearly uniform appearance, tracing the mid-to-low density structures that interface between voids and higher-density regions.
    \item Filaments: Display a highly anisotropic distribution, forming the most recognizable visual feature of the cosmic web by connecting regions of higher density, albeit with some distortion and dilution in the SDSS dataset due to redshift space distortions.
    \item Knots: Manifest as concentrated regions at the intersections of filaments, representing the highest density areas.
\end{itemize}

Our algorithm introduces an innovation compared to previous work: the classification of random points. 
We therefore present visual results for this previously unexplored dataset type in the context of the cosmic web.

Figure \ref{classesALLRandom} provides a complementary perspective, showing the classification of random points for the same slice as in Figure \ref{classesALL}. 
This figure has several notable features:

\begin{itemize}
    \item The presence of voids is now clearly discernible, contrasting sharply with the filamentary structures.
    \item The distribution of random points on sheets remains spatially homogeneous across different inputs.
    \item The filaments in the random points, although they follow the overall filamentary pattern in the object catalog, are now fluffier due to the constraint of these random points having a number density close to the average.
    \item Random points classified as knots are sparse, unlike the concentrated distribution seen in the data points.
\end{itemize}

The stark differences between the classifications of object points (Figure \ref{classesALL}) and random points (Figure \ref{classesALLRandom}) underscore the ASTRA algorithm's effectiveness in distinguishing genuine cosmic structures and its potential to use the random point distribution to trace and find the underdense object distribution. 
In what follows we provide more quantitative results on these datasets.

\subsection{Classification Uncertainty}

\begin{figure}
	\includegraphics[width=1.0\columnwidth]{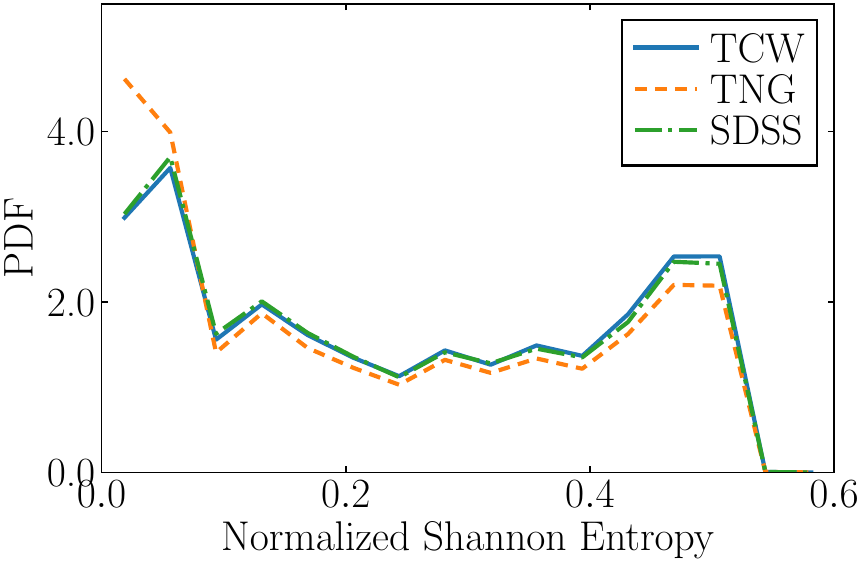}
    \caption{Probability density function of the entropy values calculated on the real points of each of the three catalogs. 
    In all cases the entropy most of the points have entropy between 0 and 0.50, which indicates that, in general, the algorithm decides between two environments.}
    \label{entropyALL}
\end{figure}

The robustness of ASTRA's classification is demonstrated by the entropy values, Eq. (\ref{eq:entropy}), typically falling between 0 and 0.5, as shown in Figure \ref{entropyALL}. 
This figure presents the probability density function (PDF) for the entropy of points in all three catalogs computed using 100 different random catalogs. 
We find that only 0.05\% of the points show entropy values larger than 0.5, with the largest entropy values being 0.56.

The confined range of entropy values indicates that the classification algorithm typically decides between at most two environments for each point, suggesting that the underlying $r$ values (measuring relative connectivity to object versus random points) remain relatively stable across different random catalogs. 
These small variations in $r$ only cause points to shift across classification thresholds between adjacent cosmic web types, rather than producing dramatic changes in classification.

\begin{figure}
	\includegraphics[width=1.0\columnwidth]{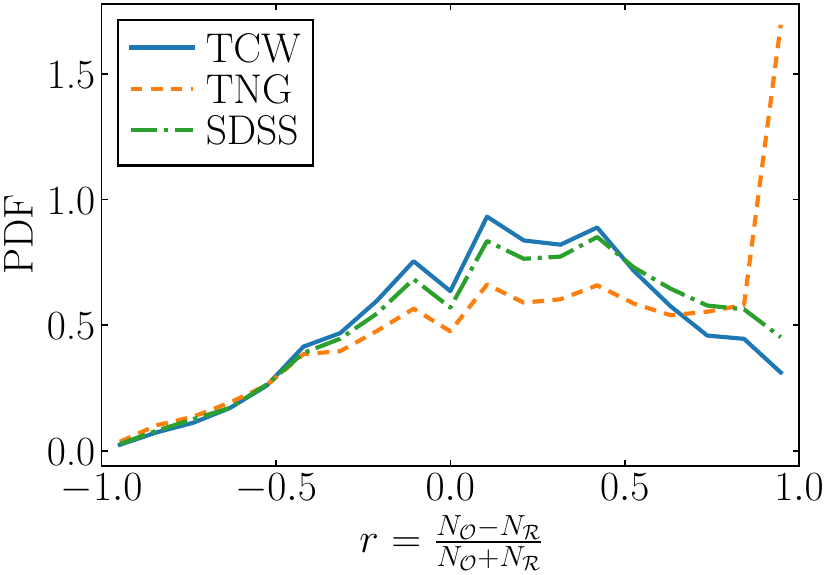}
    \caption{Probability density function of the $r$ values on the real points of each of the three catalogs.}
    \label{rvaluesALL}
\end{figure}

Further evidence of classification stability is provided in Figure \ref{rvaluesALL}, which displays the estimated PDF for the $r$ parameter across our three datasets. 
A striking feature is the prominent peak in the last bin, corresponding to values of $r = 1$, which represents points exclusively connected to other data points. 
This can be explained by different selection effects: in TNG, it reflects substructure within the most massive halos (since TCW contains only halos, not subhalos), while in SDSS it reflects the limited number of satellite galaxies around the bright galaxies in the catalog.

The distributions also reveal notable similarities between the TCW and SDSS catalogs, a pattern that is mirrored in their entropy histograms (Figure \ref{entropyALL}). 
The $r$ values exhibit an increasing trend in their histograms until reaching a peak around $r = 0.25$ in regions classified as filaments, after which the trend reverses.

This probabilistic approach offers significant advantages for cosmic web studies, providing both classifications and quantitative measures of classification reliability. 
The stability of the classification, evidenced by the confined entropy range and consistent $r$ value patterns, suggests that ASTRA's results are robust and reliable.
The ability to quantify classification uncertainty through the entropy measure makes ASTRA particularly valuable for studies where understanding the reliability of cosmic web classification is important for subsequent analyses.

\begin{figure}
	\includegraphics[width=1.0\columnwidth]{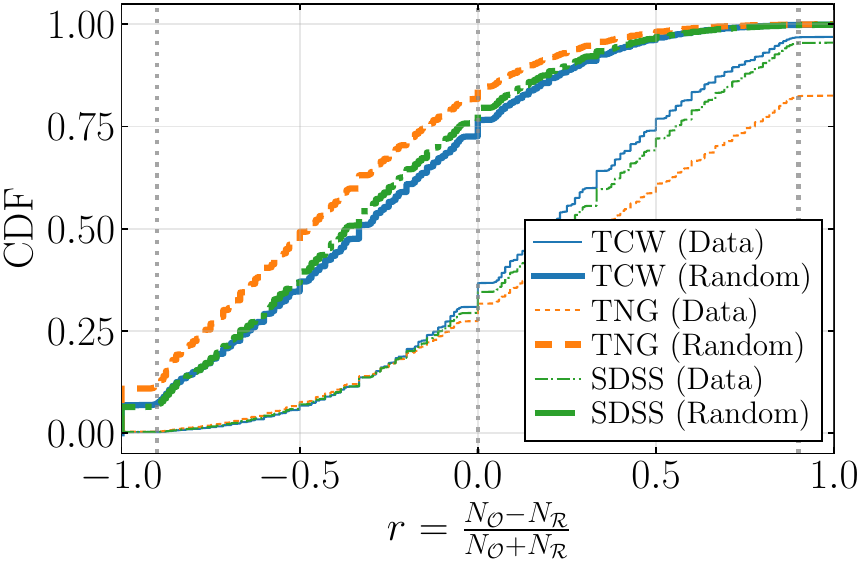}
    \caption{Cumulative probability distribution of the $r$ values on the object and random points of each of the three catalogs. 
    From this plot one can estimate different threholds if one wanted to have different volume filling fractions for the cosmic web types.}
    \label{rvaluescum}
\end{figure}

\begin{table*}
\caption{Count and mass fraction of each structure in each of the three catalogs. For the count fraction we also compute it over the random point catalog and use it as an estimate of the volume filling fraction of each environment. 
We compute the mean value and the standard deviation over 100 Monte Carlo iterations of ASTRA.}
\centering
\begin{tabular}{|l|c|c|c|c|}
\hline \hline
& \multicolumn{4}{c}{Count fraction}    \\ \hline
Catalog    & Voids   & Sheets   & Filaments   &  Knots  \\ \hline
TCW$_{\text{obj}}$     &  (0.12 $\pm$ 0.03)\%  & (35.72 $\pm$ 0.98)\%  & (62.42 $\pm$ 0.69)\%    &   (1.74 $\pm$ 0.36)\%  \\ 
TNG$_{\text{obj}}$     & (0.22 $\pm$ 0.02)\%   & (31.45 $\pm$ 0.82)\%    & (51.41 $\pm$ 1.40)\%   & (16.91 $\pm$ 2.03)\%    \\ 
SDSS$_{\text{obj}}$     &  (0.14 $\pm$ 0.04)\%  & (33.89 $\pm$ 0.98)\%   & (63.58 $\pm$ 0.63)\%    &  (2.38 $\pm$ 0.46)\%  \\ \\
TCW$_{\text{rand}}$     &  (7.28 $\pm$ 0.05)\%  & (69.28 $\pm$ 0.09)\%   & (23.30 $\pm$ 0.07)\%   &    (0.14 $\pm$ 0.01)\%  \\ 
TNG$_{\text{rand}}$     & (11.54 $\pm$ 0.09)\% & (73.13 $\pm$ 0.11)\% & (15.26 $\pm$ 0.07)\%   & (0.07 $\pm$ 0.01)\%  \\ 
SDSS$_{\text{rand}}$     & (6.78 $\pm$ 0.09)\% &  (72.60 $\pm$ 0.13)\%  &  (20.45 $\pm$ 0.08)\%  & (0.17 $\pm$ 0.01)\%    \\ \hline \hline
& \multicolumn{4}{c}{Mass fraction} \\ \hline
TCW (dark matter) &  (0.020 $\pm$ 0.001)\% & (15.00 $\pm$ 1.08)\%   & (70.38 $\pm$ 3.15)\%   & (14.60 $\pm$ 3.80)\% \\
TNG (stellar)& (0.08 $\pm$ 0.01)\%   & (20.07 $\pm$ 0.71)\%   & (44.92 $\pm$ 2.37)\%  & (34.93 $\pm$ 2.99)\% \\ \hline
\end{tabular}

\label{table:fraction}

\end{table*}

\begin{figure*}
	\includegraphics[width=2.0\columnwidth]{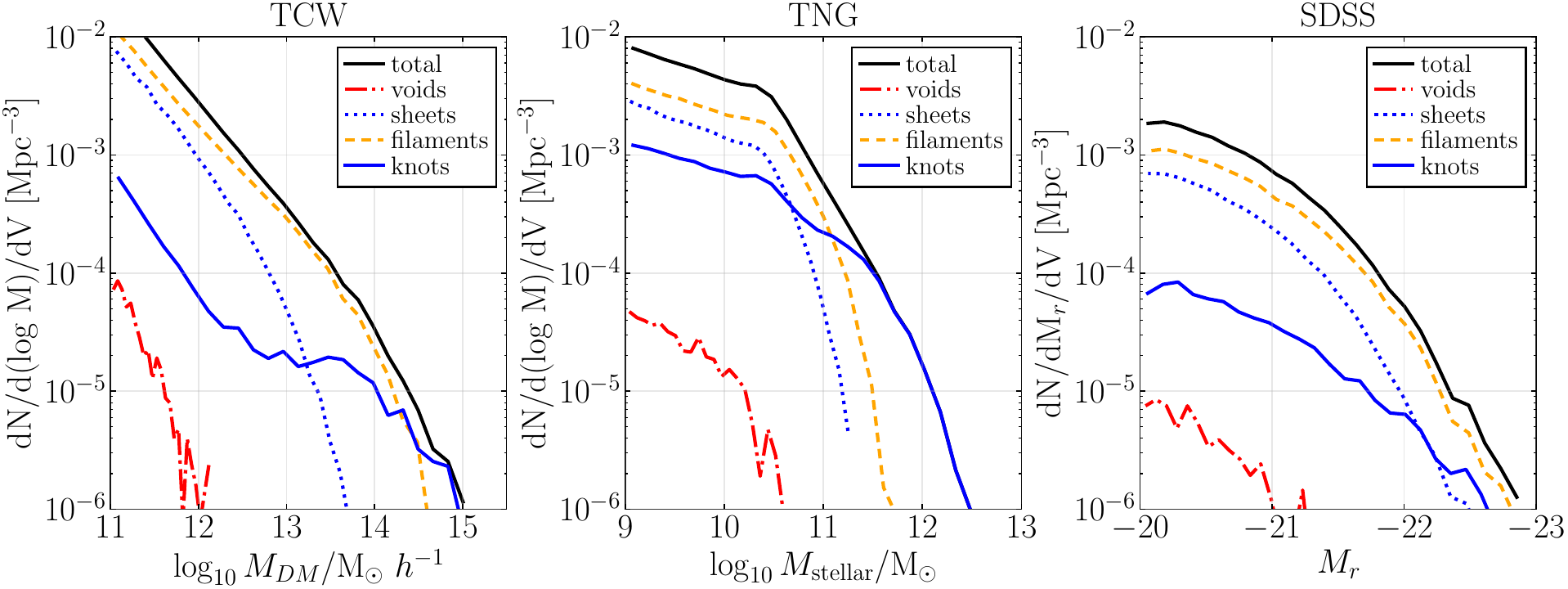}
    \caption{Different mass and luminosity functions for the different cosmic web environments. 
    From left to right: dark matter halo mass on TCW, stellar mass on TNG and luminosity on SDSS.}
    \label{fig:massdistributions}
\end{figure*}

\subsection{Volume Filling Fractions and Mass Segregation}

We compute the fraction of points (data and random) that are found in each of the web types and then calculate the mean value and standard deviation for this fraction from 100 Monte Carlo iterations. 
The results are summarized in Table \ref{table:fraction}.

We use the fraction computed on the random points as an estimate of the volume filling fraction (VFF). 
This is a reasonable approximation because the random points are designed to uniformly sample the entire survey volume or simulation box, following the same selection function as the observational data. 
Unlike galaxies, which preferentially trace overdense regions, random points provide an unbiased sampling of the entire volume. 
When ASTRA classifies these randomly distributed points into cosmic web types, the fraction of random points in each environment follows the total volume occupied by that type of environment.

These results are consistent across all three simulations, showing that in decreasing VFF values, we have: sheets, filaments, voids, and knots, with ranges between 69-73\%, 15-23\%, 6-11\%, and 0.05-0.2\%, respectively.

Comparing these results with other methods from the TCW paper \citep{TCW}, we find notable differences in VFF values. As shown in Figure 5 and Table 2 of \cite{TCW}, most cosmic web classification methods report void VFF values ranging from 40\% to 70\%. 
For instance, NEXUS identifies void volumes of 65\% and T-web reports 43\%. DisPerSE, the only other method in TCW that classifies cosmic web types from sparse data, reports VFF values of 23.9\%, 37.3\%, and 38.8\% for filaments, sheets, and voids, respectively. 
ASTRA finds substantially smaller void volumes (7\% versus 38.8\%) and larger sheet volumes (70\% versus 37.3\%).

This discrepancy in void VFF arises from our conservative threshold choice for void classification in Table \ref{c1}. 
ASTRA's current threshold ($r \leq -0.9$) is designed to identify only the most underdense void cores where galaxies are extremely rare, whereas other methods include void regions of lesser underdensity. 
Interestingly, the combined VFF for sheets and voids is remarkably similar between methods (77\% for ASTRA versus 75-80\% for most methods in \cite{TCW}), suggesting that both approaches identify similar total volumes of underdense regions but partition them differently between sheets and voids.

ASTRA's threshold can be adjusted to match other methods' void definitions. 
By changing the void classification threshold from $r \leq -0.9$ to a less stringent value, we can increase the void VFF to match typical values in the literature. 
Figure \ref{rvaluescum} shows the cumulative probability distribution of $r$ values for both object and random points across our three catalogs, providing a reference for estimating how different threshold choices would affect volume filling fractions. 
For instance, a less conservative void threshold of $r \leq -0.5$ would yield void VFFs in the range of 35\% to 50\%, much closer to values reported by other methods. 
Similar flexibility exists in other approaches, such as the T-web algorithm \citep{T-web}, where adjusting eigenvalue thresholds for the Hessian of the gravitational potential produces VFF values of 16\%, 60\%, 24\% and 1\% for voids, sheets, filaments and knots, respectively. 
This ability to tune classification boundaries allows researchers to adapt ASTRA for different scientific objectives, whether focusing on extremely underdense void cores or capturing broader underdense regions.

For the TCW and TNG simulations, we are able to compute the fraction of dark matter mass and stellar mass found in each structure, respectively. 
The relative ranking of mass fractions is consistent across the simulations. 
Most of the mass is found in filaments, followed by sheets, then knots, and finally voids. 
The percentages across the simulations vary. 
In the case of the DM simulation, almost 70\% of the mass is found in filaments, while only 45\% of the stellar mass is found in the same environments. 
This trend is inverted for knots, where 14\% of the DM mass is found in haloes in these environments, whereas up to 34\% of the stellar mass is found in the same type of environment.

The substantial difference in knot count fraction between TCW (1.74\%) and TNG (16.91\%) reflects the underlying data structure. 
TCW contains only FOF halos without substructure, while TNG includes both halos and subhalos. 
ASTRA identifies individual subhalos within groups and clusters as knots, resulting in a higher knot fraction.
This explains why, despite having ten times more objects classified as knots, the TNG knots contain about twice the mass fraction (34.93\%) compared to TCW knots (14.60\%).

\begin{figure*}
	\includegraphics[width=1.8\columnwidth]{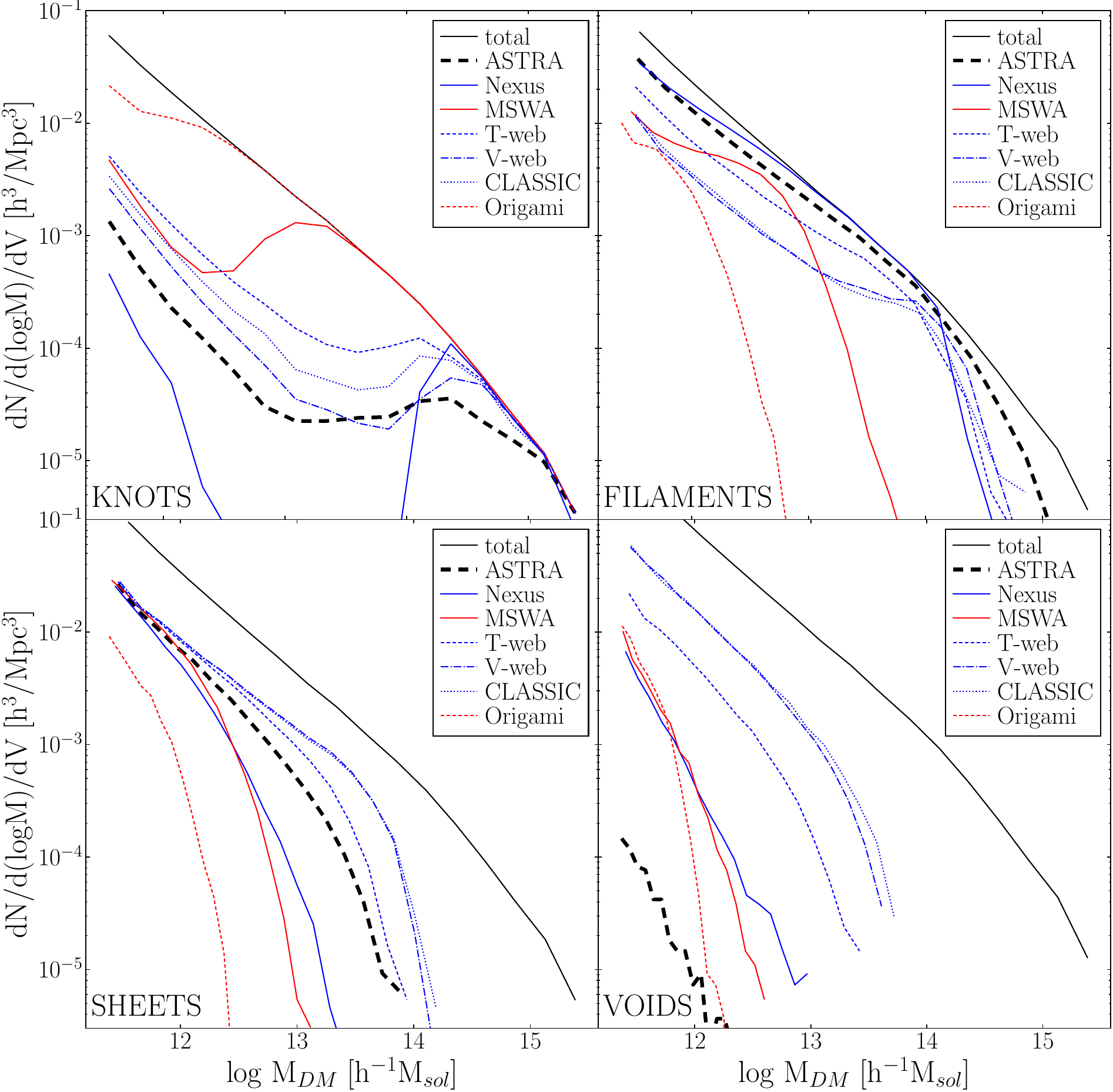}
    \caption{Mass functions comparing the results by ASTRA to the methods used in the TCW comparison project.}
    \label{fig:massdistributionsTCW}
\end{figure*}

The agreement between TCW and SDSS web fractions, despite their different physical nature, arises from similar sampling strategies. 
Both effectively represent one object per halo—TCW by containing only FOF halos, and SDSS through its bright magnitude limit that preferentially selects central galaxies. 
On the other hand, TNG's inclusion of both centrals and satellites creates a more complete but structurally different representation of the cosmic web, particularly in dense regions.

Taking the ratio between the mass fraction and the volume fraction, one can achieve a mass density estimate (in units of the average mass density), which for the DM halo mass in TCW yields: $2\times 10^{-3}$, 0.2, 3.0, and 100 for voids, sheets, filaments, and knots, respectively. 
For the stellar mass in TNG, it yields $6\times 10^{-3}$, 0.3, 2.9, and 500 for the same environments. 
While this density ordering is inherent to our method, these mass density values should not be interpreted as validation of the classification but rather as a characterization of the mass distribution within the number-density-defined environments.

To provide a broader picture, we show in Figure \ref{fig:massdistributions} different mass and luminosity functions split across environments. The main trends in these distributions are:
\begin{enumerate}
\item Objects classified as being in voids consistently show the lowest masses/luminosities.
\item Most of the objects are found in filaments, spanning all the mass/luminosity range. The exception is the objects from simulations, where the most massive systems are exclusively located in knots.
\item Sheets follow the mass/luminosity distribution of filaments, but their abundance is consistently lower than in filaments.
\end{enumerate}

The clear difference in the distribution of the brightest/most massive objects between SDSS (filaments) and simulations (knots) can be attributed to the impact of redshift space distortions (RSD) in the observational data \citep{2019MNRAS.485.5276F}. 
Without RSD corrections, the "Fingers of God" effect stretches dense clusters along the line of sight, causing many bright central galaxies physically located in knots to appear displaced into filament regions in redshift space. 
This effect, combined with the bright magnitude limit of our SDSS sample, explains why the brightest galaxies are classified as filament objects despite similar count fractions between SDSS and TCW. 
In simulations where we have access to real-space positions, this effect is absent.

Figure \ref{fig:massdistributionsTCW} compares ASTRA's halo mass functions for each cosmic web type with those from other methods in the TCW comparison project. 
We do not find a severe disagreement with the VFF results. 
The mass functions for voids are the lowest among all methods, with good correspondence to our smaller VFF values. 
For filaments and sheets, our results are similar to the T-web method, while for knots we show closer agreement with the V-web algorithm. 
The hierarchy among all the mass functions across methods follows the expected density trends (voids < sheets < filaments < knots), which are the physically relevant characteristics of cosmic web classification. 

This demonstrates that ASTRA successfully captures the fundamental density hierarchy of cosmic structures, consistent with other established methods, while placing slightly different boundaries between environments. 
The consistency in mass function shapes across methods, despite differences in absolute counts, validates ASTRA's approach to cosmic web classification based on the local relationship between data and random points.

For the purposes of this introductory paper, we've presented the native observables for each dataset (dark matter mass for TCW, stellar mass for TNG, and $r$-band magnitude for SDSS). 
However, we recognize that computing $M_r$ magnitudes for TNG galaxies would enable more direct comparison with observational SDSS data. 
In future work, we plan to implement this approach to directly compare the luminosity-environment relationship between simulations and observations, which would help disentangle physical environmental effects from observational biases such as redshift space distortions. T
his would be particularly valuable for quantifying how RSD affects the distribution of bright galaxies across different cosmic web environments.

\begin{figure*}
	\includegraphics[width=0.65\columnwidth]{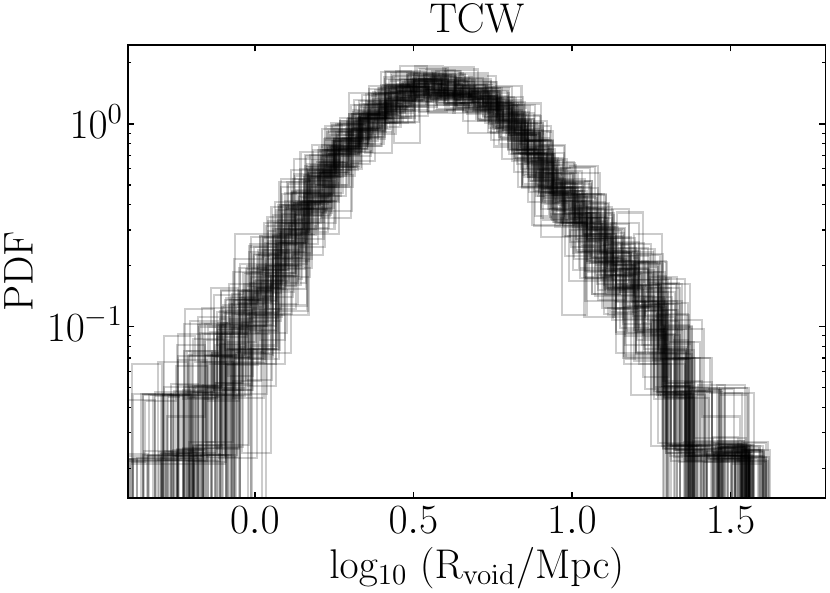}
	\includegraphics[width=0.65\columnwidth]{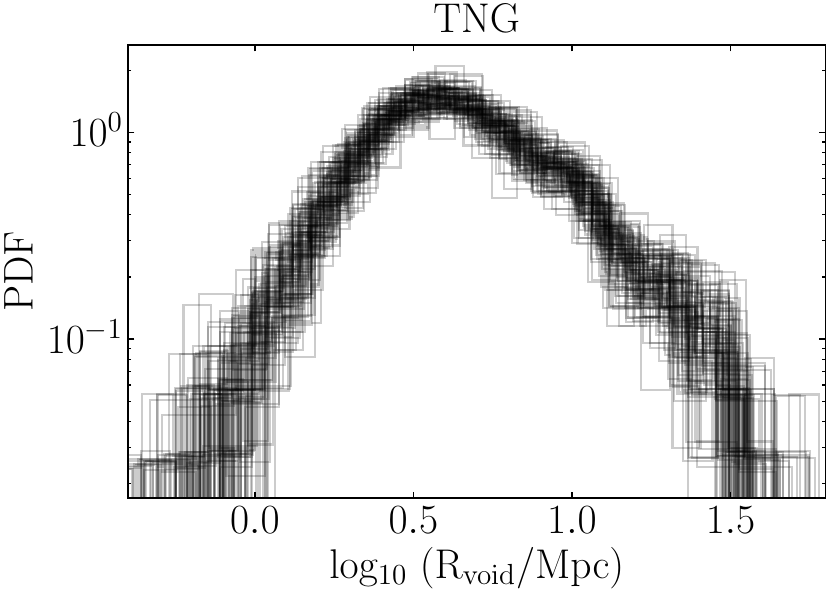}
	\includegraphics[width=0.65\columnwidth]{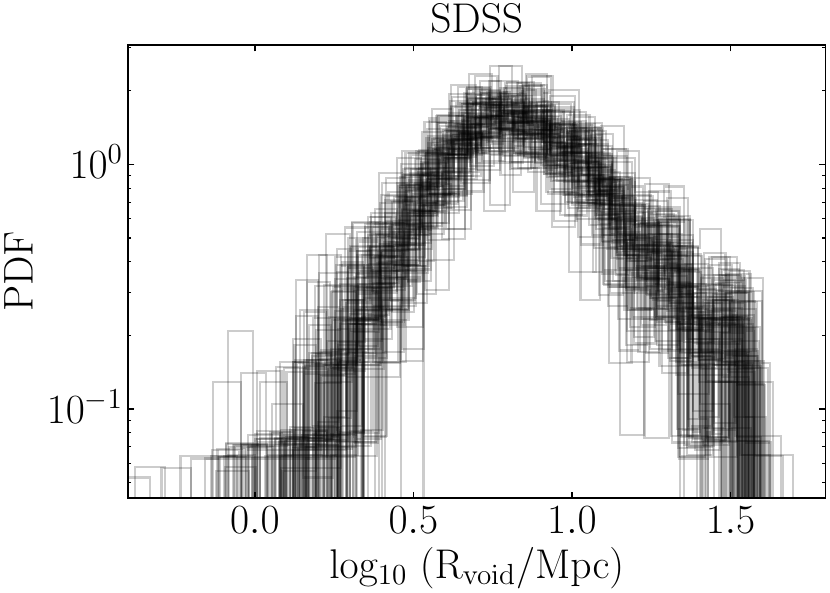}
    \caption{Void size functions. 
    There are 100 different lines in each panel, each line corresponds to a different iteration of the ASTRA algorithm. 
    Each panel, from left to right, corresponds to the TCW, TNG and SDSS datasets. 
    We keep voids that are traced with at least four random points. 
    The resulting void size distributions follow theoretical expectations \citep{ShethVoids} and align with results from other void finding algorithms. 
    ASTRA's conservative void definition (using $r \leq -0.9$) identifies smaller void volumes (approximately 7\% VFF) compared to other methods (typically 40-60\% VFF), focusing on the most underdense core regions.
    The maximum void radii are around 30 Mpc, smaller than the 40-75 Mpc maximum reported by other void finders applied to similar data \citep{SDSSVOID}.}
    \label{fig:voidsizepdf}
\end{figure*}

\subsection{Void Catalogs}

The main innovation of ASTRA compared to other cosmic web identification methods is its ability to classify random points into cosmic web environments. 
This capability is particularly valuable for void identification, as it enables us to trace underdense regions that are poorly sampled by actual galaxies. 
By detecting connected random points classified as voids in the Delaunay graph, ASTRA generates void catalogs in each iteration. 
For each identified void, we compute various geometric properties including the number of constituent points, the inertia tensor, and its eigenvalues. 
Since cosmic voids typically exhibit non-spherical morphologies, we estimate the effective void radius $R_{\text{void}}$ as the square root of the average of the three inertia tensor eigenvalues.

Figure \ref{fig:voidsizepdf} presents the probability distribution function of void radii (in logarithmic scale), including only voids traced by at least 4 random points. 
The resulting void size distributions follow theoretical expectations \citep{ShethVoids} and align with results from other void finding algorithms \citep{ShandarinVoids}. 
It is worth noting that different void finding methods applied to the same dataset can yield substantially different results. 
Recent analyses of SDSS DR7 data \citep{SDSSVOID} show that void volume filling fractions can range from 40\% to 60\% depending on the algorithm used (compared to 7\% in our case), with largest void radii between 40-75 Mpc (versus 30 Mpc maximum in our implementation). 
These differences highlight ASTRA's more conservative void definition, focusing on the most underdense core regions. A complete comparison with different void finding techniques and parameterizations is planned for future work.

The viability of ASTRA's voids for statistical applications is supported by several key results: (1) visual inspection confirming these regions correspond to genuine underdense areas (Figures \ref{classesALL} and \ref{classesALLRandom}),  (2) mass segregation patterns showing the expected lowest-mass galaxies in voids (Figure \ref{fig:massdistributions}), and (3)  the void size function follows theoretical expectations (Figure \ref{fig:voidsizepdf}).
These results demonstrate that while ASTRA identifies smaller void volumes, these regions represent physically meaningful void cores suitable for statistical analysis.

Voids have also been extensively studied and utilized to constrain cosmological parameters by using the void size function and spatial cross-correlation between void centers and galaxies \citep{VoidsDE, VoidsBOSS, VoidsBossDR12}. 
However, the precision of measurements utilizing spatial cross-correlation is limited by a numerical challenge: while the number of galaxies used could be on the order of $10^6$, the number of void positions for cross-correlation is typically on the order of $10^3$. 
This disparity of three orders of magnitude introduces noise into the numerical estimation of cross-correlation, involving the construction of a 2D histogram of relative distances for approximately $10^9$ galaxy-void pairs.

In our approach, we utilize random points that sample voids, not just their centers. With the number of random points matching the number of galaxies, and approximately 10\% of random points tracing voids, the galaxy sample of $10^6$ can be cross-correlated with $10^5$ positions representing voids. 
This increase of two orders of magnitude in the number of galaxy-void pairs used for the 2D cross-correlation function enhances precision. 
While a thorough examination of this method's potential for constraining cosmological parameters is reserved for future work, we explore in the next section its feasibility by analyzing the results of measuring the auto and cross-correlation functions for different web types in both data and random samples, as measured by ASTRA.

\subsection{2-point cross correlations}

\begin{figure*}
	\includegraphics[width=2\columnwidth]{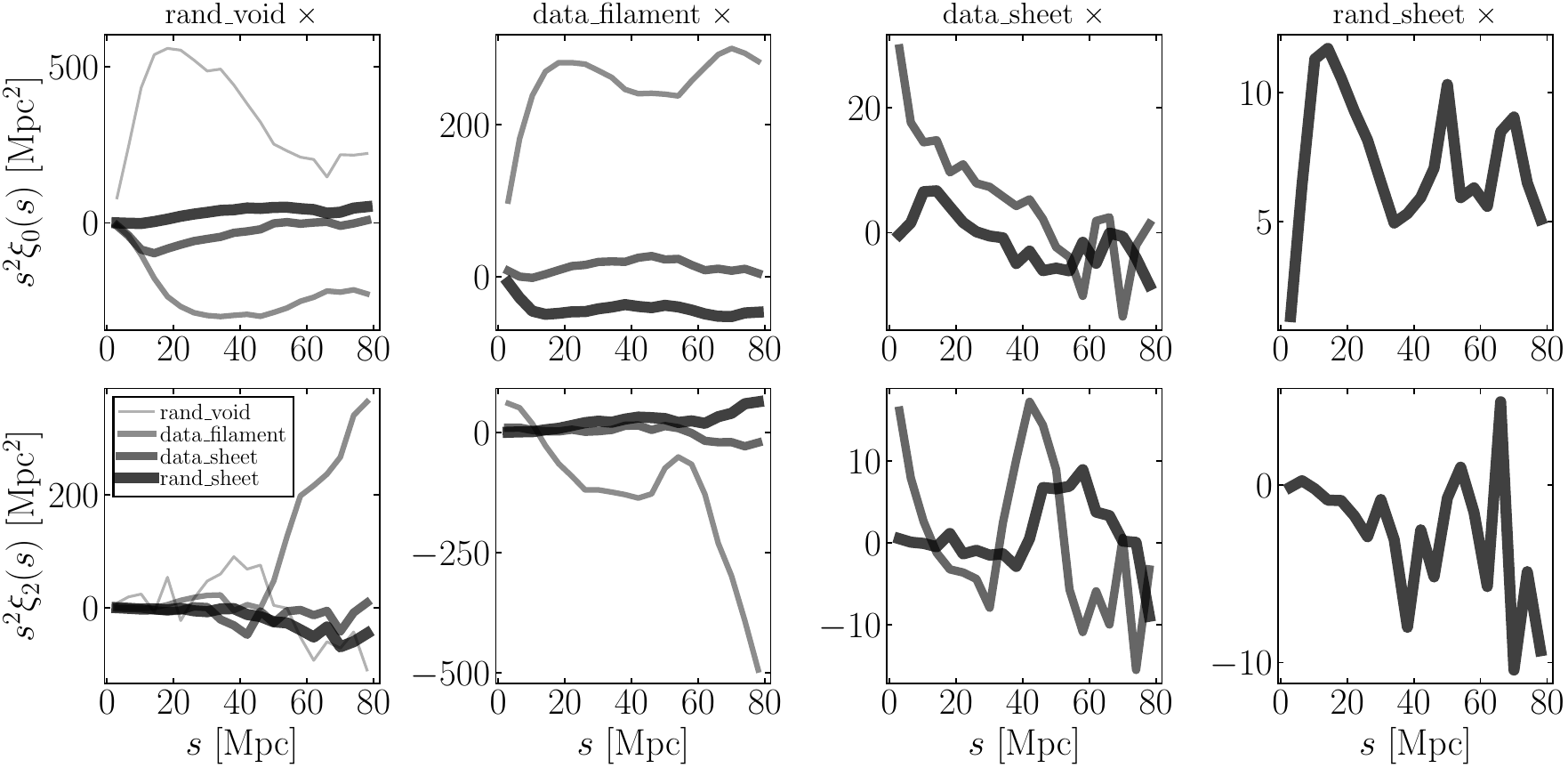}
    \caption{Two-point correlation function multipoles measured using the SDSS catalogue. 
    Each column represents a different point set (random void, data filament, data sheet and random sheet), while the inset displays the corresponding dataset against which the correlation functions are computed. 
    We observe from this analysis that the most prominent signals arise from the auto-correlations of random voids and data filaments, along with their respective cross-correlations.}
    \label{fig:correlations}
\end{figure*}

The clustering of galaxies in each web-type can be quantified using the 2-point correlation function (2PCF), which characterizes the excess probability of finding a galaxy within a given distance of another galaxy compared to a random distribution \citep{1983ApJ...267..465D}. 
In the observed universe, the galaxy distribution appears anisotropic due to radial velocity perturbations introduced by peculiar velocities of galaxies. 
This motivates the use of two coordinates, $s$ and $\mu$, to describe galaxy separations, where $s$ represents the pair separation along the line of sight, and $\mu$ is the cosine of the angle between the vector connecting the galaxies and the observer's line of sight.

A commonly used estimator for the 2PCF \citep{1993ApJ...412...64L} involves a random point distribution as a reference and is described as:
\begin{equation}
\xi(s,\mu) = \frac{DD - 2DR + RR}{RR},
\end{equation}
where $DD$ is the number of galaxy pairs in the $s,\mu$ bin, $DR$ is the number of galaxy-random pairs, and $RR$ is the number of random-random pairs. 

To differentiate different angular components in the 2PCF, the function can be projected into Legendre polynomials to compute the multipole moments defined by:
\begin{equation}
\xi_\ell(s) = \frac{2\ell + 1}{2}\int_{-1}^{1} \xi(s,\mu) P_{\ell}(\mu) d\mu.
\end{equation}

In this paper, we focus on the monopole, $\xi_0(s)$, and the quadrupole, $\xi_{2}(s)$, measured on the volume-limited data from SDSS presented in previous sections. 
No weights are applied to galaxies to account for observational systematics, and no Feldman-Kaiser-Peacock weights \citep{1994ApJ...426...23F} are used to correct for variations in the number density of galaxies. 
This simplification was chosen deliberately as implementing proper weighting schemes requires detailed characterization of the survey's imaging and spectroscopic systematics, which extends beyond the scope of this introductory method paper. 
While weights would improve the precision of absolute correlation amplitudes and enable more direct comparison with literature values, our primary focus is on the relative patterns across different cosmic web environments and their characteristic scales, which remain qualitatively robust even without weights. 
Future applications of ASTRA aimed at precise cosmological constraints will incorporate appropriate weighting schemes.

The correlation functions are measured in 61 $\mu$ bins from -1 to 1 and 21 radial bins from 0.1 to 80 Mpc. Randoms are used with 10 times more points than the original galaxy catalog.
These additional random points are only used for the correlation function estimation and do not affect the cosmic web classification.

We also measure cross-correlations between two different samples, $D_1$ and $D_2$, using:
\begin{equation}
\xi(s,\mu) = \frac{D_1D_2 - D_1R_2 - D_2R_1 + R_1R_2}{R_1R_2},
\end{equation}
where $R_1$ and $R_2$ are two different random sets.

In our case, we use four different samples: data-sheets, data-filaments, random-sheets, and random-voids, corresponding to samples in their respective datasets. 
We avoid using samples with a low number of points such as data-voids, data-knots, and random-knots. 
These datasets allow us to compute six different cross-correlations and four different self-correlation functions.

The main results for these correlations are shown in Figure \ref{fig:correlations}. 
The monopole exhibits its largest amplitudes for the auto-correlation of data-filaments, random-voids, and their cross-correlation. 
Furthermore, these correlations show a distinctive transitional scale around $20$ Mpc. 
The next high-amplitude cross-correlation is found for data-sheets and random-voids, also with a transitional scale around $15$-$20$ Mpc. 
Some of the quadrupoles show large anisotropies, the largest being found for the auto-correlation of data-filaments and their cross-correlation with random-voids, especially for scales larger than $40$ Mpc. 
This might be due to the fact that filaments and voids are expected to show large redshift space distortions which influence their selection effect when identified in redshift space.

We anticipate that a data vector composed by the concatenation of all the self and cross-correlation results could be used to constrain cosmological parameters \citep{2023MNRAS.522..606P}. 
This involves a complex process that includes predicting the expected covariance matrix for all observables, taking into account observational biases, instrumental limitations, and efficiently exploring the cosmological parameter space. 
Such a process is beyond the scope of this paper and is left for future work.

\begin{table*}
\caption{Overview of the methods used in the \emph{Tracing the Cosmic Web} \citep{TCW}. 
The Input distinguishes between dense (tipycally DM computational particles from an N-body simulation) and sparse (typically DM haloes or galaxies).
}
\centering
\begin{tabular}{|l|c||c|c|c|cc|}
\hline \hline
Method    & Web types & Input      & Grid based & Main Reference & F1-score\\ \hline
\textbf{ASTRA}  & all  & sparse  & No & \textbf{This paper} & 1.0 \\ 
MSWA  & all  & dense  & No &\cite{MSWA}  & 0.55 \\ 
T-web  & all  & dense  & Yes & \cite{T-web}  & 0.59 \\ 
V-web  & all  & dense  & Yes & \cite{V-web}  & 0.36\\ 
CLASSIC  & all  & dense  & Yes &\cite{CLASSIC}  & 0.37\\ 
NEXUS +  & all  & dense &  Yes &\cite{NEXUS}&  0.65\\ 
ORIGAMI  & all  & dense & No &\cite{Origami}&  0.35 \\ 
DisPerSE  & all except knots  & sparse & No &\cite{disperse} &  0.68\\
SpineWeb  & all except knots  & dense  & Yes &\cite{spineweb}&  0.55\\ 
MMF-2  & all except knots  & dense & Yes & \cite{MMF-2}&  0.64\\ 
Bisous  & filaments  & sparse  & No & \cite{BisousPaper} &  0.41\\ 
FINE  & filaments  & sparse  & Yes & \cite{FINE}&  0.69\\ 
 \hline
\end{tabular}
\label{tab:all_methods}    
\end{table*}

\begin{figure*}
	\includegraphics[width=2\columnwidth]{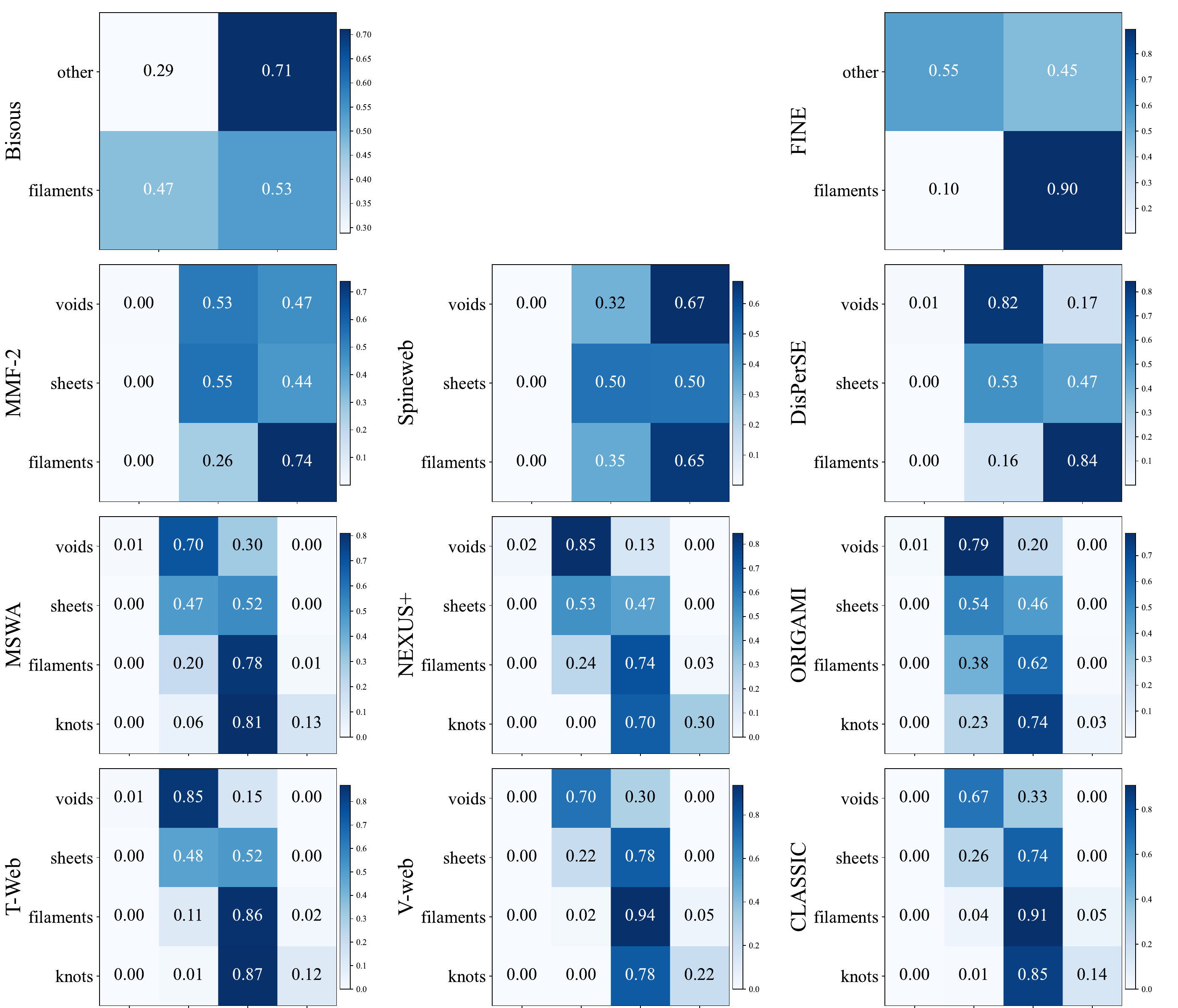}
    \caption{Confusion matrices built by comparing the results given by ASTRA on the TCW catalog against those given by the methods cited in Table \ref{tab:all_methods}. 
    The first row corresponds to the methods that can only classify dark matter halos in filaments, the second those that can classify them in all environments except knots, and the last two rows correspond to the remaining 6, which can, like ASTRA, classify the points in each of the four structures.}
    \label{CMsALL}
\end{figure*}

\subsection{Comparison Against Other Cosmic Web Finding Methods}

With the aim of gaining deeper insight into ASTRA's capabilities, we present a comparative analysis against other cosmic web identification methods.
Our goal is to identify the method that most closely resembles ASTRA's results based on the TCW simulation, incorporating findings from various cosmic web detectors.

Table \ref{tab:all_methods} provides an overview of the methods employed in the TCW project, for which public data exists on the classification of FOF DM haloes. 
The table outlines the web types each method can categorize, the types of input data it can handle (dense, from simulations, or sparse, from observations), and whether it operates on a grid-based system. 
To assess the similarity between the results obtained using different methods and those obtained using ASTRA, we computed confusion matrices and calculated the weighted average F1 score across different structures.
The F1 score is calculated as the harmonic mean of precision and recall, where precision denotes the ratio of true positive results to the total number of positive results found, and recall represents the ratio of true positive results to the total number of results that should have been classified as positive.

Figure \ref{CMsALL} presents the confusion matrices for the 11 methods compared to the results of all methods in the TCW paper \citep{TCW}. 
Two key observations emerge from this figure. 
Firstly, the most significant discrepancies in classification occur for voids, where ASTRA often classifies DM halos as sheets, contrary to other methods. 
Secondly, the highest level of agreement in classification generally occurs for filaments. 
The classification accuracy across the six methods that categorize into four structures typically follows a decreasing order: filaments, sheets, knots, and voids.

The last column in Table \ref{tab:all_methods} summarizes the F1 score for each comparison. 
Among methods producing a four-type classification, NEXUS yields the most similar results to ASTRA, with an F1 score of 0.65. 
For three-type classification, the highest F1 score is obtained in the comparison with DisPerSE, scoring 0.68. 
When comparing against filament classifiers, the best result is achieved in the comparison with FINE, yielding an F1 score of 0.69. Moreover, the highest level of agreement across methods and cosmic web environments is observed for filaments in the V-web algorithm, with 94\% of the galaxies classified as filaments also identified by ASTRA.

From this comparison, we conclude that ASTRA stands out among other methods for its unique ability to classify all web types, handle both dense and sparse data points, and operate without requiring a grid. 
Furthermore, the haloes classified as filaments appear to be the most agreed-upon result when compared against other methods. 

\section{Conclusions}

The large-scale distribution of galaxies in the Universe reveals a complex network known as the cosmic web, which is conventionally classified into four primary morphological types: voids, sheets, filaments, and knots. 
In this work, we have introduced ASTRA, a new classification method that uses both observational data and randomly distributed points to provide a complete view of cosmic structures.

ASTRA's approach differs from previous methods by directly incorporating random points in the classification process. 
This approach offers specific advantages for identifying underdense regions that are typically undersampled by galaxies, while still effectively characterizing overdense structures. 
The method works on sparse 3D point distributions without requiring density field interpolation or fixed geometric assumptions, making it computationally efficient and applicable across diverse datasets.

Our application of ASTRA to three distinct datasets (a dark matter-only simulation, a hydrodynamical simulation, and SDSS observational data) yields several insights:

\begin{itemize}
\item Visual and statistical analyses confirm that ASTRA successfully identifies expected cosmic web patterns, with knots concentrated at filament intersections, filaments forming connecting networks, sheets appearing as planar structures, and voids representing underdense regions.

\item Volume filling fractions derived from random points show a consistent pattern across datasets, with sheets occupying the largest volume (69-73\%), followed by filaments (15-23\%), voids (6-11\%), and knots (0.05-0.2\%). 
Our void identification focuses on extremely underdense cores, resulting in smaller void volumes compared to other methods but capturing physically meaningful regions.

\item Mass and luminosity distributions show the expected environmental segregation, with the lowest-mass objects predominantly found in voids and the highest-mass objects in filaments and knots. 
The classification of bright SDSS galaxies into filaments rather than knots likely reflects the impact of redshift space distortions, which remain an important consideration when applying ASTRA to observational data without appropriate corrections.

\item The method enables construction of void catalogs with size distributions consistent with theoretical expectations, offering potential applications for cosmological parameter constraints through void statistics.

\item Correlation function analyses reveal significant signals in both auto and cross-correlations between different cosmic web types, highlighting distinct transitional scales around 15-20 Mpc that warrant further investigation.
\end{itemize}

While ASTRA shows promise, we acknowledge several limitations that need attention in future work. 
The method currently uses conservative threshold values ($r \leq -0.9$ for voids) that identify only the most extreme void cores, differing from broader void definitions in some existing methods.
However, this represents a design choice rather than a limitation: the $r$ parameter framework allows users to adjust thresholds for their specific research questions. 
For instance, relaxing the void threshold would increase void volume filling fractions to match some literature values, while different filament/sheet boundaries could highlight various structural features.

Application to observational data without redshift space distortion corrections affects our classification, particularly for dense structures. 
As with all cosmic web classifiers, there is a specific choice of connection pattern to be made (Delaunay tessellation in our case), though our tests indicate this choice provides a solid foundation for environmental classification.

Despite these limitations, our comparative analysis with other cosmic web finding techniques shows that ASTRA provides classifications that align with established methods, particularly for filamentary structures. 
The method's main strengths (working with sparse data, handling both overdense and underdense regions, and avoiding smoothing or interpolation) make it a useful addition to existing cosmic web analysis tools. 
The method's ability to create detailed void catalogs and measure cross-correlations between different cosmic web elements may improve cosmological parameter estimation and expand our understanding of structure formation in the Universe.

Going forward, we expect that ASTRA's approach to cosmic web classification, particularly its identification of various environment types and use of random points for underdense region mapping, will be useful for current and upcoming large spectroscopic surveys.
In particular, we plan to apply our algorithm to the data from the Dark Energy Spectroscopic Instrument (DESI), taking particular interest into the Large Scale Structure catalogs provided by the collaboration \citep{DESILSS}, which already include carefully constructed randoms as required by ASTRA.

\section*{ACKNOWLEDGEMENTS}
We thank our reviewers, including Miguel Aragon-Calvo, for their valuable insights and specific advice that greatly improved this manuscript. 

JEFR acknowledges funding from Facultad de Ciencias at Universidad de los Andes under project INV-2023-162-2846. 

\section*{CONFLICT OF INTEREST}
Authors declare no conflict of interest.

\section*{Data Availability Statement}
The datasets analyzed in this study are publicly available from the following sources:

TCW simulation data: The Tracing the Cosmic Web simulation catalog is available through the original comparison project. Details for data access can be found in \citep{TCW} and associated project documentation.

TNG simulation data: The Illustris-TNG simulation data (TNG300-1) used in this work is publicly available through the IllustrisTNG project website (\url{https://www.tng-project.org/data/}). Galaxy catalogs can be accessed via the TNG web-based API or downloaded directly.

SDSS observational data: The Sloan Digital Sky Survey Data Release 7 galaxy catalog is publicly available through the SDSS website (\url{https://www.sdss.org/}). The NYU Value-Added Galaxy Catalog used in this analysis can be accessed at \url{http://sdss.physics.nyu.edu/vagc/}.

ASTRA implementation: The ASTRA algorithm implementation, example scripts, and analysis code used to generate the results in this paper are available at: \url{https://github.com/forero/DelaunayASTRA}



\bibliographystyle{rasti}
\bibliography{refs} 








\bsp	
\label{lastpage}
\end{document}